\documentstyle[12pt,a4]{article}

\makeatletter
\@addtoreset{equation}{section}

\makeatother

\def\ga{\alpha} \def\gb{\beta}
\def\gG{\Gamma}
\def\gd{\delta} \def\gD{\Delta} \def\ge{\epsilon}
\def\gw{\omega} \def\gW{\Omega} \def\gK{\chi}
\def\gfi{\phi}  \def\gr{\rho}   \def\gvfi{\varphi} 
\def\gm{\mu}   \def\gn{\nu}  \def\gs{\sigma} \def\gS{\Sigma} \def\gt{\tau}
\def\gq{\theta} \def\gl{\lambda} \def\gL{\Lambda} 
\def\n{\noindent}
\def\np{Nucl. Phys. {\bf B}}
\def\pl{Phys. Lett. {\bf B}}

\def\ijmp{Int. J. Mod. Phys. {\bf A}}
\def\cmp{Comm. Math. Phys.}
\def\prd{Phys. Rev. {\bf D}}
\def\be{\begin{equation}}  \def\ee{\end{equation}}   \def\dag{\dagger}
\def\it{\item}
\def\CC{\kern 0.27em \vrule height1.45ex width0.03em depth0em \kern-0.30em%
\rm C}
\newcommand{\ignore}[1]{}

\def\tG{\tilde G}
\def\tB{\tilde B}
\def\tD{\tilde D}

\begin{document}

\hoffset = -1truecm
\voffset = -2truecm

\title
{\bf GRAVITATIONAL INSTANTONS AND BLACK PLANE SOLUTIONS IN 4-D STRING
THEORY }
\vskip 2cm
\author{
ADRIAN R. LUGO\thanks{E-mail: LUGO@ICTP.TRIESTE.IT}\\
\normalsize International Center for Theoretical Physics \\
\normalsize Strada Costiera 11, (34100) Trieste, Italy
}

\date{November 1994}
\vskip 2cm

\maketitle
\begin{abstract}

We consider gauged Wess-Zumino models based on the non compact
group $SU(2,1)$.
It is shown that by vector gauging the maximal compact subgroup $U(2)$ the
resulting backgrounds obey the gravity-dilaton one loop string vacuum
equations of motion in four dimensional euclidean space.
The torsionless solution is then interpreted as a pseudo-instanton of the
$d=4$ Liouville theory coupled to gravity.
The presence of a traslational isometry in the model allows to get another
string vacuum backgrounds by using target duality that we identify with
those corresponding to the axial gauging.
We also compute the exact backgrounds.
Depending on the value of $k$, they may be interpreted as instantons
connecting a highly singular big bang like universe with a static singular or
regular black plane  geometry.

\end{abstract}

\newpage


\section{Introduction}

Since Witten's discover [1] that singular solutions to the string vacuum
equations of motion [2] can be represented by exact two dimensional
conformal field theories known as gauged Wess-Zumino-Novikov-Witten models
(GWZM) [3], a lot of work has been made in the last years about the subject,
with special attention put on solutions of relevance in black hole
physics and cosmology [4].
In $1+1$ dimensions the ``famous" $SU(1,1)/U(1)$ coset representing a
Schwarzchild like black hole has been  exhaustively analyzed.
Generalizations of this model as $SO(d-1,2)/SO(d-1,1)$ cosets
were considered in [5,6], where a guess leading to the exact
(to all orders in $1/k$) backgrounds was given.

Of course we are ultimately interested in realistic four dimensional
models.
Some of them, obtained essentially by taking tensor products of
$SU(1,1)$'s and $U(1)$'s, were considered in [7,8].
A possible classification of cosets leading to effective target spaces with
one time direction was given in [9].

In this paper we consider a model based on gauging the maximal compact
subgroup $U(2)$ of the non compact group $SU(2,1)$.
The interest is at least two-fold.
First, the backgrounds by themselves represent a highly non trivial solution
to the string equations, or matter coupled to gravity system; from general
arguments the one loop solution should have euclidean signature and then
represent some kind of gravitational instanton, but this view could be
changed by considering the exact solution.
Second, we think it is an instructive algebraic exercise to explicitely
work out non abelian groups other than those related to the $A_1$ Lie
algebra. The techniques, in particular the parametrizations, used here
for $SU(2,1)$ are in principle extensive to general $U(p,q)$.

The paper is organized as follows.
In Section 2 we set up general definitions and conventions, while
Section 3 is devoted to the $SU(2,1)$ parametrizations.
In Section 4 we describe the computation of the one loop effective
backgrounds, in Section 5 the curvature and equations satisfied by them.
In Section 6 we compute the (presumibly) exact backgrounds and conjecture
possible interpretations.
In Section 7 we quote the expressions of the one loop dual solution.
Section 8 is devoted to the conclusions.
An appendix divided in three sections is added, where we collect some
useful formulae.

\newpage


\section{Conventions}

A bosonic string that sweeps out an euclidean genus $g$ world-sheet $\gS$
embedded in a gravity-axion-dilaton $d$ dimensional background on target
space ${\cal M} $ is described by the action
\be
S[X;G,B,D] = \frac{k}{4\pi}
           \int_{\Sigma} \, \left(\, (G_{ab} (X) *\, +\, i B_{ab} (X) ) dX^a
\wedge  dX^b  - \frac{1}{2 k} D(X)*R^{(2)} \,\right)
\ee
where ``*" stands for the Hodge mapping wrt some metric on $\gS$, $R^{(2)}$
being its Ricci scalar that satisfies
$\;\;  \int_{\gS}*R^{(2)} = 8 \pi ( 1 - g ) \;\;$.

The Weyl invariance condition of this two dimensional sigma model imposes
that, at one loop
\footnote{ Strickly speaking, at first order in $\frac{1}{k} \equiv \ga '$,
see i.e. [10]. }
the backgrounds satisfy the set of equations [2]
\begin{eqnarray}
0 &=& R_{ab} - \, \nabla_a \nabla_b D  - \frac{1}{4} H_{acd} H_{b}{}^{cd} \cr
0 &=& -2\; \nabla_a \nabla^a D -  \nabla_a D \nabla^a D + R
- \frac{1}{12} H_{abc} H^{abc} + \gL\cr
0 &=& \nabla^{c}( e^{D} H_{abc})
\end{eqnarray}
where $\; H\equiv dB\;$ and $\;\gL=\frac{26-d}{3} k\,$
(our definitions for curvature, ecc., are those of ref. [11]).
These equations follow from the $d$-dimensional action on ${\cal M}$
\be
I[G,B,D] = \int_{\cal M} \, e^{D}\; (*R +  \,\nabla D \wedge *\nabla D
         -  \frac{1}{12} H \wedge *H + *\gL)
\ee

A GWZM is defined as follows.
Let $G$ a Lie group, $H$ a subgroup of $G$ and ${\cal G}, {\cal H}$ their
respective Lie algebras.
If $g: \Sigma \mapsto G$,
$\gw (g) \equiv g^{-1}dg = - {\overline \gw}(g^{-1}) \in {\cal G}$
stand for the Maurer -Cartan forms, and ${\cal A} \in {\cal H}$ is a gauge
connection, then the defining action of a GWZM is [3]
\begin{eqnarray}
S[g,{\cal A}]
&\equiv& \frac{k}{4\pi} ( I_{WZ}[g] + I_{G}[g,{\cal A}] )
\equiv \frac{k}{4\pi} (  I_{0}[g] + i \gG [g] + I_{G}[g,{\cal A}] ) \cr
I_{WZ} [g] &=& \frac{1}{2} \int_{\Sigma} tr( \gw (g) \wedge *\gw
(g))  +
 \frac{i}{3} \int_{ {\cal B}, \partial {\cal B}= \Sigma }
tr( \gw (g) \wedge \gw (g) \wedge \gw (g) ) \cr
I_{G} [g,{\cal A}] &=&
\int_\Sigma tr \big( - {\cal A} \wedge (* + i1) \,\gw(g)
                \; +\; {\cal A} \wedge (* - i1) \,{\overline \gw} (g) \cr
&-& g \, {\cal A}\; g^{-1} \wedge (* +i1) {\cal A} \; +\; {\cal A} \wedge *
{\cal A} \big)
\end{eqnarray}
where $``tr"$ is normalized in such a way that the lenght of a long rooth
of $\cal{G}$ is $2$ [12].
\ignore{
\footnote{ For $A_n$ algebras the trace in the fundamental representation
works.}
}
This action is invariant under the gauge transformations
\footnote{A slightly modified version of the action (2.4) and
gauge transformations (2.5), the so called ``axial" gauging, is possible
if $\cal{H}$ contains abelian subalgebras;  the effective target is
different but both theories are equivalent (dual), in agreement with
current algebra arguments.}
\newpage
\begin{eqnarray}
g^{h}&=&h \, g \, h^{-1} \cr
{\cal A}^{h}&=&h \, {\cal A} \,h^{-1} - {\overline\gw}(h)
\end{eqnarray}
for an arbitrary map $\;h : \gS\mapsto H \,$.

\noindent If we pick a basis $\,\{T_a,\, a=1,\ldots, {\rm dim} H \}$ in
$\cal H$, then by integrating out the gauge fields in $I_G$ we obtain the one
loop order effective action
\begin{eqnarray}
   S_{eff}[g] &=& \frac{k}{4\pi} W[g] - \frac{1}{8\pi} \int_\Sigma D(g)
*R^{(2)} \cr
         W[g] &=& I_{WZ}[g] + {\tilde I}[g] \cr
{\tilde I}[g] &=& -2 \int_\Sigma\, \frac{1}{l} ({\gl^c})^{ab} \,
                  a_a\wedge(*-i1) b_b
\end{eqnarray}
where $l=l(g)$ and $\gl^c =\gl^c(g)$ are the determinant and the cofactor
matrix of
\be
\gl_{ab}(g) = \frac{1}{2}\; tr( T_a T_b - g T_a g^{-1} T_b)
\ee
and
\begin{eqnarray}
i\, 2\, a_a = tr( T_a \,\gw(g)) \cr
i\, 2\, b_a = tr( T_a \,{\overline \gw(g)})
\end{eqnarray}
Clearly the gauge invariance condition $\;S_{eff}[g^h]=S_{eff}[g] \;$
makes the effective target dependent on $\; d= {\rm dim}G-{\rm dim}H\;$
gauge invariant field variables constructed from $g$.
The $d$ dimensional metric and torsion are read from $W[g]$.
The dilaton field
\be
D(g) = \ln | l(g)|
\ee
comes from the determinant in the gaussian integration leading to (2.6) after
convenient regularization [9].
\footnote{Because $\gl$ transforms as a 2-tensor in the adjoint representation
of $H$, $D(g)$ will be gauge invariant for subgroups with semisimple Lie
algebra; for non semisimple subalgebras action (2.6) does not exist,
see below.}

It is undoubtly of major importance to get $d=4$ target spaces since they
can represent realistic backgrounds for string theory, with implications
in cosmology and black hole physics in particular.
Models with one time direction have been clasified in [9].
Most of them consist of groups product of $SU(1,1)'s $ and $U(1)'s$ (see
however [5], where the only ``less" trivial SO(3,1)/SO(2,1) coset is briefly
considered).
Unfortunately one of the most interesting targets, the ``stringy"
Schwarzchild solution (and more generically, geometries with a high degree
of isometries), has evaded us.
A naive explanation of this fact could be the following one: since at one
loop $R_{ab} = 0$ and $D=const.$ for this solution, we would have to have
(up to $g$-independent normalizations) $l(g)=1$.
But from (2.7) we see that the $\gl$ matrix is null when we approach to
$g=1$, and certainly it cannot have a non-vanishing determinant.
More generally speaking, if $G$ is semisimple we can always choose an
orthogonal set of generators in $\cal G$ of non-zero norm; if we write $g=T
U$ with $U\in H$ and $T\in G/H$ then from (2.7) we get
\be
\gl (g) = (1 - S(T) R(U))^t h
\ee
where $R(U)$ is the adjoint representation matrix of $U$, $S(T)$ contains
the adjoint action of the coset element $T$ on the $\cal H$ generators and
$h$ is the Killing-Cartan metric on $\cal H$.
For elements in $ H (S(1)=1) $, $ \gl$ becomes singular on some
submanifold (the target space nature of it to be elucidated) and
$l(g) \neq 1$.
If $G$ is not semisimple, then the Killing-Cartan form
has null eigenvalues and $\lambda $ does not exists in general.
In any case is hard to see how a singular target space (and, to one loop at
least, it should be!) could raise with a constant dilaton in the present
context of GWZM.
Maybe the non abelian duality transformations recently introduced [13,14]
could indirectly lead to an exact conformal field theory
representation of the stringy (and others) Schwarzchild black hole.
\footnote{We point out that the S-duality [15] recently introduced
in the context of superstring theory does not hold here; in fact our
solutions has non zero cosmological constant.}

\newpage

\section{The SU(2,1)/U(2) model}

Coming back to our problem, a certainly non trivial four dimensional target we
would  get by considering $G=SU(2,1)$ and $H=U(2)$.
{}From general arguments it will have (at one loop!) euclidean signature [9],
and so it could represent some kind of ``gravitational instanton" in the
general sense of reference [16].
\footnote{By means of a Wick rotation we can get $(++--)$ signature; it
corresponds to gauging the $U(1,1)$ subgroup.}
So let us concentrate on this model.
In view of the gauge invariance of the theory, it will be of most
importance to fix a convenient parametrization.
We will denote vectors with bold-type letters; matrices will be understood
from the context.

An arbitrary element $g \in SU(2,1)$ admits the coset decomposition wrt
its maximal compact subgroup $U(2)$,
\be
g = T({\bf c}) H(U,u^* )
\ee
where $T,H$ are given in eqns. (A.4).
Clearly the SU(2,1) topology is $\Re ^4 \times S^3 \times S^1$.
Now, outside the origin of $\CC ^2$ the complex 2-vector ${\bf c}$ can be
uniquely written  as ${\bf c}=s\, {\bf n}$, with
$ s\equiv ({\bf c}^\dag {\bf c})^\frac{1}{2} $
being the radial coordinate of $\Re ^4$ and ${\bf n}^{\dag} {\bf n} = 1$.
The unitary vector ${\bf n}$ is in one-to-one correspondence with a $SU(2)$
matrix
\be
{\bf n} \equiv \left( \matrix{
                        n_1 \cr
                        n_2 \cr } \right)
\Leftrightarrow
N\equiv \left( \matrix{
                           n_1^*   &n_2^* \cr
                          -n_2     &n_1   \cr } \right)
\ee
Since an arbitrary element of $U(2)$ can be written as
\be
 U=\left(
            \matrix{ u &0 \cr
                     0 &1} \right) P
\ee
with $P \in SU(2)\,$ and $u\equiv e^{i\gvfi}=detU \,$, we can parametrize
the $ U \in U(2) $ in (3.1) as
\be
 U = N^{\dag} \left(
              \matrix{ u &0 \cr
                     0 &1 \cr}\right) P \, N
\ee

\n and then we rewrite $g$ in the form
\footnote{ From now on we will use the variable $t \in [0,\infty )$
and the symbols $\; s\equiv \sinh t \; ,\; c\equiv \cosh t $. }
\be
         g = H(N^\dag ,1)\, e^{t \gl_4}\,e^{i \frac{\gvfi}{2}(\gl_3 +
\sqrt{3}\gl_8)} H(P,1)\, H(N,1)
\ee
where the relations $ N\,{\bf n} = \left( \matrix{ 1\cr
                                             0\cr} \right)$ and (A.6) were
used.
Finally, if according to (C.3) we introduce
\newpage
\begin{eqnarray}
  X &\equiv& e^{i \frac{\gvfi}{2} \gs_3 } \, P =
e^{i \frac{\gq}{2} \gs_3} \; {\overline X} \; e^{-i \frac{\gq}{2}\gs_3} \cr
  V &\equiv& e^{i \frac{\gq}{2} (1 - \gs_3 ) } N  \;\; \in U(2)
\end{eqnarray}
we obtain
\be
g = H(V^{\dag} ,1) \, e^{t \gl_4}\, e^{i \frac{\sqrt{3}}{2} \gvfi\gl_8 } \,
H({\overline X},1)  H(V,1)
\ee
It is clear from this parametrization that $V$ is a gauge variable and
decouple from the model.
The remaining four gauge invariant variables (for example,
$ (t,\gvfi, x_0, x_3) ) \;$ locally parametrize the effective target
manifold whose topology might be naively identify with
$\Re ^2\times {\cal D}\,$ where $\cal D$ is a disk.
This can be seen from the fact that according to (3.1,4) and (A.6), the
(complex) variables
\begin{eqnarray}
{\bf c}^\dag \, N^\dag \, U\, N \, {\bf c} &=&
s^2 \, (x_0 + i x_3) \, e^{i\frac{\gvfi}{2}} = s^2 \, (p_0 + i p_3) u  \cr
tr U &=& 2\, x_0 \, e^{i\frac{\gvfi}{2}} = p_0 (1 + u) - i \, (1-u) \,p_3
\end{eqnarray}
are the gauge invariant ones, and belongs to $\Re ^2 $ and $\cal D$
respectively.
\footnote{The complex variable $trU$ (that encodes $det\, U = u $) is the
gauge invariant variable describing the coset $U(2)/Adj\, U(2)\equiv {\cal
D}$. We thank M. Blau for a discussion on this point.}
However as follows from (2.10), the origin of $\Re ^2$ as well as the
boundary of the disk will become singular.

We remark that $X$ belongs to $SU(2)$ only ``locally" , but not
globally as $P$ does; it rises from parametrizing a $U(2)$ matrix as a
$SU(2) \times U(1)$ element in (3.3), $ U = e^{i \frac{\gvfi}{2} } X $.
It is useful to carry out computations and we will also consider it in what
follows, as well as with
\begin{eqnarray}
V &=& e^{i \frac{\gfi}{2} } \; (v_0 1 + i \, {\bf v} \cdot {\bf \gs} ) \cr
1 &=& v_0{}^2 + {\bf v}\cdot {\bf v}
\end{eqnarray}
in Section 6.

\newpage


\section{Computation of the one loop metric}

In this section we will describe with some detail the calculations of the one
loop  backgrounds.
The parametrization (3.7) (with $V=1$) will be assumed.

First of all, we have to choose a convenient basis in $\cal H$.
We take the following generators ($\; (\check{e}_i )_j = \gd_{ij} \;$)
\begin{eqnarray}
T_i &=& \gl_i - \frac{1}{\sqrt{3}} \,\gl_8 \;\gd_{i,3}\; ,\;\;\; i=1,2,3\cr
T_4 &=& -\frac{2}{\sqrt{3}} \,\gl_8
\end{eqnarray}
In the notation of Appendix B, we compute from (2.7) the matrix $\gl$ to be
\begin{eqnarray}
  M &=& 1 - R^t A \, ,       \;\;\;\;\; A = c\, 1 - (c-1) Q \cr
{\bf m_1} &=& s^2 \, (R^t - 1) \, \check{e}_3 \cr
{\bf m_2} &=& {\bf 0} \cr
m_0 &=& - 2 \, s^2
\end{eqnarray}
where $R\equiv R(X)$ is given in (C.7) and $\; Q=\check{e}_3
\check{e}_3{}^t $.

Now from (B.2) we get
\be
\gl^c = \left( \matrix{
                         m_0 M^c      &0\cr
                     -{\bf m_1}{}^t M^c     &m\cr} \right)
\ee
where (c.f. (B.4) )
\begin{eqnarray}
M^c&=&(1 + (c-1) R_{33} - c\, trR ) \, 1 + c R + c R^t
      + c(c-1)\, R^t Q - (c-1) Q R \cr
m&=&- s^2 (1 - R_{33} )
\end{eqnarray}
The next step is to compute the vectors in (2.8).
They are given by
\begin{eqnarray}
  {\bf a}&=&{\bf U} - \frac{1}{2} d\varphi \, \check{e}_3 \cr
a_4&=& - d\varphi \cr
  {\bf b}&=&A \; {\overline {\bf U}} - \frac{1}{2} d\varphi \,
\check{e}_3 \cr
b_4&=& -( 1 + \frac{3}{2} s^2 )\, d\varphi - s^2\, {\overline U}_3
\end{eqnarray}
On the other hand, the Wess-Zumino action (2.4) results
\be
I_{WZ}[g] = \int_\gS \, (dt \wedge *dt - \frac{3}{4} d\varphi\wedge
* d \varphi ) + I_{WZ}[X]
\ee
With (4.3,6) and after some calculations we get (2.6) in the form
\be
W[g] = \int_\gS \big( dt\wedge * dt + \frac{1}{s^2  (1 -R_{33})}\,
       ( L_{\varphi\varphi} d\varphi\wedge *d\varphi + L_{XX} -
         L_{X\varphi} )  \big) + i \gG [X]
\ee
where ( $S\equiv (1- c \, trR)\, 1 + c \, R^t + c^2 R $ )
\begin{eqnarray}
L_{\gvfi\gvfi}&=&\frac{1}{2} (R M^c)_{33} +
                 \frac{1}{4} (1-R_{33}) (1 + 3 c^2 ) \cr
L_{XX}&=&- s^2 (1 - R_{33}) \, {\bf U}\cdot \wedge * {\bf U} +
          2 \, {\bf U}\cdot \wedge (* - i1) M^c \, A \,{\overline {\bf U}} \cr
L_{X\gvfi}&=& \check{e}_3 \cdot S \, {\bf U}  \wedge (* - i1) d\gvfi +
     {\overline {\bf U}} \cdot S \, \check{e}_3 \wedge (* + i1) d\gvfi
\end{eqnarray}
and after repeatedly using formulae collected in Appendix C we get
\begin{eqnarray}
L_{\gvfi\gvfi}&=&-1+ R_{33} + \frac{1}{4} (5-3R_{33}) (c-1)^2 +
                 \frac{c}{2}\, (5-2R_{33}-trR)\cr
L_{XX} &=& 2\, (c-1)^2  dx_3\wedge * dx_3 +
           2\, (c+1)^2 dx_0 \wedge * dx_0 \cr
       &+& i 2 s^2 (1 - R_{33}) (x_0 dx_3 - x_3dx_0) \wedge d\gq \cr
L_{X\gvfi}&=&2\, d\gvfi \wedge * ( s^2 x_0\, dx_3 -  (s+2)^2 x_3\, dx_0 )
+ i s^2 (1-R_{33}) \, d\gq \wedge d\gvfi \cr
\gG[X]&=& -2\,\int_\gS \, (x_0 \, dx_3 - x_3 \, dx_0)\wedge d\gq
\end{eqnarray}
{}From these results we learn two important facts:
\begin{itemize}
\item the last term in $L_{XX}$ cancels the $\gG$ contribution;
\item the last term in $L_{X\gvfi}$ drops out because it gives a total
derivative contribution to $W$;
\end{itemize}
that lead us to conclude that:
\begin{enumerate}
\item the $\gq$ variable in $X$ decouples, as should.
As we saw in Section 3 this is only a non trivial check of gauge invariance;
\item the three terms that cancel are those that could give rise to the
axionic field $B$, in other words the target obtained is $torsionless$.
\end{enumerate}
This last fact is not expected ``a priori".
To our kwowledge, a classification of torsionless groups in GWZM is not
available.
{}From the model considered here we can argue that the key fact for this to
happen lies in the possibility of going to a gauge in which the Wess-Zumino
term is zero
\footnote{In WZM we certainly have zero torsion if $\gG = 0$.}
(which is made explicit in 1.), but a more general argument is
lacking.

If the backgrounds are defined as in (2.1), we read from (4.7,9) the
non-zero    metric components in the $(t,\gvfi , x_0, x_3)$ variables
( $ \gr \equiv +\, \sqrt{1 - x_0{}^2 - x_3{}^2}$ )
\begin{eqnarray}
G_{tt}&=&1\cr
G_{\gvfi\gvfi}&=&\frac{c^2}{s^2} +
\frac{c-1}{4(c+1)} \frac{x_0{}^2}{\gr ^2 } +
\frac{c+1}{4(c-1)} \frac{x_3{}^2}{\gr ^2} \cr
G_{00}&=&\frac{c+1}{c-1} \frac{1}{\gr ^2} \cr
G_{33}&=&\frac{c-1}{c+1} \frac{1}{\gr ^2} \cr
G_{0\gvfi}&=&  \frac{c+1}{2(c-1)} \frac{x_3}{\gr ^2 } \cr
G_{3\gvfi}&=& -\frac{c-1}{2(c+1)} \frac{x_0}{\gr ^2 }
\end{eqnarray}
and from (2.9), (B.3) and (4.2,4) the dilaton field
\be
D = \ln( s^4 \gr ^2 ) + D_0
\ee
We notice here the existence of a manifest isometry, a traslation in the
$\gvfi$ variable with Killing vector $K_\gvfi = \partial_\gvfi\;$.

If we go back to $P$ variables (3.6) by means of the rotation
($0\leq R \leq \pi /2 \,$)
\ignore{
\be
 \left( \matrix{x_3 \cr
                 x_0 \cr} \right) = e^{i \frac{\gvfi}{2} \gs_2 }
  \left( \matrix{p_3 \cr
                 p_0 \cr} \right)
\ee
}
\be
x_0 + i\, x_3 = (p_0 + i\, p_3 ) \, e^{i\frac{\gvfi}{2}} = \sin R \,
e^{i\psi}  = \sin R \, e^{i (\psi_P + \frac{\gvfi}{2} )}
\ee
the metric takes the form
\footnote{ $2\, dx\, dy \equiv dx\otimes dy + dy\otimes dx $.}
\begin{eqnarray}
G &=& dt^2 \, +\, \frac{c^2}{s^2} \,\, d\gvfi ^2\, +\,
\frac{1}{s^2 \, \gr ^2} \,
( \; | c \; e^{i\gvfi} + 1 |^2 \, dp_0{}^2
 + | c \; e^{i\gvfi} - 1 |^2 \, dp_3{}^2 \cr
&-& 4\, c \sin\gvfi \; dp_0 \, dp_3 \; ) \cr
&=& dt^2 \, +\, \frac{c^2}{s^2} \,\, d\gvfi ^2\, +\,
\frac{1}{s^2} \;(\; |c\, e^{i2\psi}+1| ^2 \, dR^2
+ |c\, e^{i2\psi}-1| ^2 \, \tan ^2 R \, d \psi_P{}^2 \cr
&-& \, 4\, c\, \tan R \, \sin 2\psi \; dR\, d\psi_P \; )
\end{eqnarray}
\ignore{
where we have introduced
$\; p_0 + i p_3\equiv \sin R e^{i\psi_P}\; $ and $\; x_0 +ix_3\equiv \sin R
e^{i\psi} \; ,0\leq R\leq \frac{\pi}{2}$.
}
In this coordinates the metric looks simpler (in particular, has only
one non diagonal term), but the isometry is not manifest.

\newpage


\section{The curvature and the equations of motion.}

It is convenient in what follows to introduce an orthonormal basis
$\, \{\gw^a\} \,$ in the cotangent space of $\cal M$,
$G=\gd_{ab} \; \gw^a \otimes \gw^b $.
We choose
\begin{eqnarray}
\gw^1 &=& \frac{c}{s} \, d\gvfi \cr
\gw^2 &=& \frac{c+1}{s\, \gr } \, (dx_0 + \frac{x_3}{2} d\gvfi) \cr
\gw^3 &=& \frac{c-1}{s\, \gr } \, (dx_3 - \frac{x_0}{2} d\gvfi) \cr
\gw^4 &=& dt
\end{eqnarray}
and its dual in the tangent space $(\,\gw_a (\gw^b ) = \gd_a{}^b \, )$
\begin{eqnarray}
\gw_1 &=& \frac{s}{c}\, (\partial_\gvfi \, + \,\frac{x_0}{2} \partial_3
\, - \,\frac{x_3}{2} \partial_0 )\cr
\gw_2 &=& \frac{c-1}{s} \gr \; \partial_0
\cr
\gw_3 &=& \frac{c+1}{s} \gr \; \partial_3
\cr
\gw_4 &=& \partial_t
\end{eqnarray}
{}From the first Cartan's structure equation (torsionless condition)
\be
 T^a \equiv d\gw^a +\gw^a{}_b \wedge \gw^b = 0
\ee
\n we read the non vanishing connections
\footnote{Remember that in an orthonormal basis the metricity condition
$\,\gw^a{}_b =-\gw^b{}_a\, $ holds, as well as the general symmetry
properties: $ R_{abcd}=R_{cdab} =-R_{bacd} $ [11].}
\begin{eqnarray}
\gw^1{}_2 &=&  \gw^3{}_4 = \frac{1}{s} \,\gw^3 \cr
\gw^1{}_3 &=&-\gw^2{}_4 = \frac{1}{s} \,\gw^2 \cr
\gw^2{}_3 &=& \frac{c^2 + 1}{2\, s\, c} \, \gw^1
        \,+\, \frac{c+1}{s} \frac{x_3}{\gr}\,\gw^2
        \,-\, \frac{c-1}{s} \frac{x_0}{\gr}\,\gw^3  \cr
\gw^1{}_4 &=& -\frac{1}{sc} \, \gw^1
\end{eqnarray}
By using now the second Cartan's structure equation
\be
 \gW^a{}_b \equiv d \gw^a{}_b + \gw^a{}_c\wedge \gw^c{}_b =
\frac{1}{2} R^a{}_{bcd}\,\gw^c\wedge \gw^d
\ee
\n we read the Riemman curvature
tensor
\begin{eqnarray}
R_{1212} &=&  R_{1234} =  R_{3434} = \frac{1}{c+1} \cr
R_{1324} &=&- R_{1313} = -R_{2424} = \frac{1}{c-1} \cr
R_{1223} &=&  R_{2334} = \frac{2}{c+1}\frac{x_0}{\gr}\cr
R_{1323} &=&- R_{2324} =-\frac{2}{c-1}\frac{x_3}{\gr}\cr
R_{2323}&=&  \frac{2}{s^2} + R_{22} \cr
R_{1423} &=&-  R_{1414} = \frac{2}{s^2}
\end{eqnarray}
and contracting, the Ricci tensor $R_{ab} \equiv R_{cacb} = R_{ba}$
\begin{eqnarray}
R_{11} &=& R_{44} = - \frac{4}{s^2} \cr
R_{12} &=&-R_{34} = - \frac{2}{c-1} \frac{x_3}{\gr} \cr
R_{13} &=& R_{24} = - \frac{2}{c+1} \frac{x_0}{\gr} \cr
R_{14} &=& R_{23} = 0 \cr
R_{22} &=& R_{33} = -2\; ( 1 + \frac{2}{s^2} +
 \frac{c-1}{c+1} \frac{x_0{}^2}{\gr ^2} +
 \frac{c+1}{c-1} \frac{x_3{}^2}{\gr ^2}\; )
\end{eqnarray}
Finally, the scalar curvaure $R\equiv R_a{}^a$ is
\be
-\frac{1}{4} \, R =  1 + \frac{4}{s^2} +
 \frac{c-1}{c+1} \frac{x_0{}^2}{\gr ^2} +
 \frac{c+1}{c-1} \frac{x_3{}^2}{\gr ^2}
\ee
With these results at hand it is straightforward to verify that the
graviton-dilaton system given by equations (4.10,11) verify the
consistency equations (2.2) with $B=0$ and $\gL=12$.
We do not know if the torsion remains null at higher orders, but we
speculate that it is indeed the case.
As we anticipate, $t=0$ and $\gr = 0 $ are true singularities of the geometry,
where the parametrization (3.7) breaks down.

Here a little disgresion is in order.
The value of $\gL$ suggests that the model is conformally invariant
at one loop iff $k=\frac{18}{11}\simeq 1.64$.
On the other hand, from current algebra arguments [12]
the {\cal exact} central charge of the model is
\begin{eqnarray}
c_{SU(2,1)/U(2)} &=& c_{su(3)} - c_{su(2)} - c_{u(1)} =
8 \frac{k}{k-3} - 3 \frac{k}{k-2} - 1 \cr
&=& 4\, +\, 6\, \frac{3k-5}{(k-2)(k-3)}
\end{eqnarray}
Then imposing the cancelation against ghost contribution we obtain
the values $k_+ \simeq 3.96\;$ and $\; k_- \simeq 1.86$.
The second one is near the value obtained perturbatively at first order.
It is believed that by taking into account all loop corrections the
value of $\gL$ should lead to $k_+$ or $k_-$; however $k$ does not seem
to be big enough to assert that the perturbative theory necessarily
corresponds to $k_-$.
Moreover, in analogy with the condition that $-k = n $ be a
positive integer needed for the quantum consistency of the
compact models it is speculated that unitarity would allow only $\; k>3 $,
and if true (the subject is far from being well understood by now)
$k_+$  should be the right value to be considered.
We will take $k \in \Re^+$ for which at least the one loop path integral
seems to be well defined
[17]; see next section for more about.
As a last observation, if we consider the ``non critical" GWZM, i.e., with
a dynamical Liouville field, the allowed values of $k$ are rational:
$k_{\pm} = 4,\; \frac{13}{7}$.

In order to compare with euclidean Einstein gravity, we introduce the metric
$G^E\equiv e^{D} G$. Then the backgrounds $(G^E , D)$ are classical
solutions of the action
\be
S[G^E,D] = \int_{\cal M} \, (*R^E - \frac{1}{2} \,\nabla^E D \wedge
*\nabla^E D + *\gL e^{-D} )
\ee
which describes a Liouville field coupled to gravity in $d=4$, and may
then be interpreted as a ``pseudo-instanton" of this theory.
In fact the solution is singular at $t=0$ and $\rho = 0 \,$ as expected,
and the $R_{14}$ and $R_{23}$ components fail to be (anti)
self-dual, as usually known instantons are [18].
What is more, it is not asymptotically flat in the usual sense (at least in
the standard range of the coordinates of the model that we assume), and
gives an infinite value for the action
\be
I_{inst} = 12 \, \pi^2 \, \sinh^4 T\, e^{D_0}
\ee
where $T$ is a cut-off in the $t$-integration. In the compact coset
$SU(3)/U(2)$ the variable $t$, better to say, its continuation to
imaginary values $\, \tau \equiv i\, t \, $ is naturally bounded to the
interval $[0,\, \pi /2 ]\,$, and the action is finite.

A possible interpretation of the solution is as follows.
For $t\gg 1$ we have
\begin{eqnarray}
G &\rightarrow& \, dt^2 + d\gvfi ^2 + dR^2 + tg^2 R \, d\psi_P{}^2 \cr
D &\rightarrow& \, 4\, t + 2 \ln \cos R \cr
R &\rightarrow&  \, - 4 \, \sec ^2 R
\end{eqnarray}
which describes the topology product of a cylinder (a plane in the compact
case, for $\tau$ near $\pi /2$ ) and a ``trumpet".
On the other hand it may be thought as a euclidean coontinuation
of the non singular cosmological solution
\be
G_{cs} = -dx_0{}^2 + \tanh^2 x_0 \, dx_1{}^2\, +\, dx_2{}^2  \, +\, dx_3{}^2
\ee
arising from the $SL(2,\Re)\times SO(1,1)^2 / SO(1,1)$ model [19].
Then is tempting to interpret the instanton as a path in
``euclidean time t" that interpolates two universes, one in a ``big bang"
phase (singularity at $t=0$) and other smoothly evolving according to (5.13).
We will see in the next section that for finite $k$ very different (and
appealing) possibilities arise.

As a final remark we note that being the string coupling constant [4]
$$
g_{st} = e^{-D/2}
$$
then from (5.11,12) we have
$$
I_{inst} = \frac{3\;\pi^2}{4 \; g_{st}{}^{2}}
$$
exhibing the usual non perturbative behaviour characterizing the
``tunneling amplitud"
$\; exp(-I_{inst})$
for the process described by the instanton.

\newpage


\section{The exact backgrounds}

\n {\bf The computation}

In references [5,6] an ansatz to obtain the exact metric and
dilaton backgrounds was proposed.
Here we resume it in a few items.

\n {\bf A)} Let ${X_a}$ be a basis in $\cal G$ simple and compact,
satisfying the algebra \be
[ X_a,X_b ] = i f_{ab}{}^c X_c
\ee
and $g \in G$.
We define left and right currents (that certainly satisfy (6.1)) as linear
operators acting on $G$ according to
\begin{eqnarray}
{\hat J}_a^R g &\equiv& - g X_a \cr
{\hat J}_a^L g &\equiv& X_a \; g
\end{eqnarray}

\n {\bf B)} Once we read from (6.2) ${\hat J}_a^{L,R,}$ we construct the
quadratic  Casimir operators in this $G$-realization,
\be
{\hat \gD}_G^{L,R} \equiv g^{ab} {\hat J}_a^{L,R} {\hat J}_b^{L,R}
\ee
where $g^{ab}$ is the inverse of the Cartan metric $g_{ab} = tr(X_a X_b )$
(for normalizations, see Section 2), and in the same way we construct the
Casimir operators ${\hat \gD}_H^{L,R}$ associated with the subgroup $H$, by
restricting (6.3) to the $\cal H$ generators.
Then we define the Virasoro-Sugawara operators
\be
{\hat L}_0^{L,R} \equiv
\frac{1}{k + C_{\cal G} } {\hat \gD}_G^{L,R} - \frac{1}{k + C_{\cal H} } {\hat
\gD}_H^{L,R} \ee
where $C_{ {\cal G},{\cal H} }$ are the respective dual Coxeter numbers.
If $\cal{H}$ is semisimple then we will have sums with prefactors
corresponding to each simple components [12].

\n {\bf C)} We identify the subspace of functions on $G$ dictated by the
gauge  invariance conditions
\footnote{ We remember that $\hat{V_a} = \hat{J}_a ^L + \hat{J}_a^R$ are
the generators of the vector transformations (A.6).}
\be
({\hat J}_a^L + {\hat J}_a^R) f(g) = 0 \; , \;\;\;a=1,...,dim{\cal H}
\ee

\n {\bf D)} Finally we apply the hypothesis of [6]
\begin{eqnarray}
({\hat L}_0^L + {\hat L}_0^R ) f(g) &\equiv& - (k + C_{\cal G} )^{-1}
\gK^{-1} \partial_\gm(\gK G^{\gm\gn}\partial_\gn )f(g) \cr
\gK &=& e^D \; \sqrt{|det G|}
\end{eqnarray}
from where we can directly get $G^{\gm\gn}$ by looking at the quadratic
terms, and a system of first order differential equations to determine
$\gK$ (and so $D$) from the linear terms.

Going to our model, we take $X_a\equiv \gl_a$ the Gell-Mann matrices
and consider the parametrization (3.7,8) and (C.4).
Let us introduce the commuting linear operators
$ {\hat {\bf X}}$ and ${\hat {\bf V}}$
\begin{eqnarray}
{\hat X}_1 &=& -i\;(x_2 \, \partial_3 - x_3 \, \partial_2 ) \cr
{\hat X}_2 &=& -i\; x_0\; \partial_2 \cr
{\hat X}_3 &=& -i\; x_0\; \partial_3 \cr
V_i &=& \frac{i}{2} ( v_0 \, \partial_i - \ge_{ijk} \, v_j \, \partial_k ) ,
\;\;\; i=1,2,3
\end{eqnarray}
that verify (6.1) with $ f_{ij}{}^{k}=\ge_{ijk} $.

Then from (6.2) we read
\footnote{The index $\ga=1,2$ refers to the combinations
$\; \gl_1^{\pm} = \frac{1}{2} (\gl_4
\pm i\gl_5)\; , \;\; \gl_2^{\pm} = \frac{1}{2} (\gl_6 \pm i\gl_7) $.}

\n \underline{Right currents}
\begin{eqnarray}
{\hat R}_i &=& - R(V)_{ji}\; ( {\hat X}_j + u_j ( {\hat V}_3 - i \partial_\gfi
) ) \; ,\;\; i=1,2,3 \cr
 {\bf u} &=& \frac{1}{x_2} ( x_0 \; \check{e}_1  + x_3\; \check{e}_2 -
x_2 \; \check{e}_3 ) \cr
{\hat R}_\ga^{+} &=& - \frac{u^{3/2}}{2} (XV)_{1\ga}\; (\partial_t
+ i \frac{s}{c} \partial_\gvfi ) + A_0^V \partial_\gfi + i\; \bf{A^V} \cdot
{\bf \hat{V}} + i \, {\bf A^X} \cdot {\bf \hat{X}} \cr
i\, 2\, s\,u^{-3/2} A_0^V &=&  \frac{c^2 + 3}{2c}
(XV)_{1\ga} +
\frac{cz - z^* }{x_2}\; (XV)_{2\ga}  \cr
i\, 2\, s\, u^{-3/2}\bf{A^V} &=&  2\, (XV)_{2\ga}\; (
-\check{e}_1 + i \check{e}_2\, ) + ( \frac{s^2}{2c} (XV)_{1\ga} + \frac{cz -
z^*}{x_2}  (XV)_{2\ga} )\, \check{e}_3 \cr
i\, 2\, s\, u^{-3/2}{\bf A^X} &=& (c+1) (XV)_{2\ga} \; \check{e}_1\, -
i\, (c-1)  (XV)_{2\ga} \; \check{e}_2\, + \frac{s^2}{2c} (XV)_{1\ga} \;
\check{e}_3\cr {\hat R}_\ga^- &=& ( {\hat R}_\ga^+ )^* \cr
{\hat R}_8 &=& i \frac{2}{\sqrt{3}}\, \partial_\gvfi
\end{eqnarray}

\n \underline{Left currents}
\begin{eqnarray}
{\hat L}_i &=& -{\hat R}_i + 2 \, R(V)_{ji} {\hat V}_j \; ,\;\;\; i=1,2,3 \cr
{\hat L}_\ga^+ &=& \frac{1}{2} V_{1\ga} (\partial_t - i \frac{s}{c}
\partial_\gvfi )  + A_0^V \partial_\gfi + i \bf{A^V} \cdot \bf{\hat V} +
i \bf{A^X} \cdot \bf{\hat X} \cr
i\, 2\, s\, A_0^V &=&  -\frac{7 c^2 - 3}{2c} V_{1\ga} +
 \frac{cz^* - z}{x_2} V_{2\ga} \cr
i\, 2\, s\, \bf{A^V} &=& c\, V_{2\ga}\, (\check{e}_1 - i \, \check{e}_2 \, )
+\,
( \frac{s^2}{2c} V_{1\ga} + \frac{cz^* - z}{x_2} V_{2\ga} )\,
\check{e}_3  \cr
i\, 2\, s\, \bf{A^X} &=&  -(c+1) V_{2\ga}\, \check{e}_1 - i(c-1) V_{2\ga}
\, \check{e}_2 +  \frac{s^2}{2c}\, V_{1\ga} \, \check{e}_3  \cr
{\hat L}_\ga^- &=& ({\hat L}_\ga^+)^* \cr
{\hat L}_8 &=& - {\hat R}_8 + i \, 2\, \sqrt{3}\, \partial_\gfi
\end{eqnarray}

Clearly the first and last equations in (6.9) translate the gauge conditions
(6.5) as the independence on $\gfi$ and ${\bf v}$, i.e., on the gauge variable
$V$.
Restricting us to the gauge invariant subspace, we get the laplacians
\be
{\hat \gD}_G^L = {\hat \gD}_G^R = {\hat \gD}_{U(1)} + {\hat \gD}_{SU(2)} +
\{ {\hat R}_\ga^+, {\hat R}_\ga^- \}
\ee
and acording to (6.4) we have
\footnote{
The first equality follows from
$ \gD_G^L = \gD_G^R$ and $\gD_H^L = \gD_H^R $,
this last one valid on gauge invariant functions [5].
Also the usual change
$\; k \rightarrow -k\; $
coming from (2.4) for non compact groups is made [1].
}
\be
{\hat L}_0^L = {\hat L}_0^R = \frac{1}{k-3} {\hat \gD}_{SU(2,1)}
 - \frac{1}{k-2} {\hat \gD}_{SU(2)} - \frac{1}{k} {\hat \gD}_{U(1)}
\ee
Carrying out the computations and applying (6.6) we read the inverse
metric; the modified basis (5.1) looks
\ignore{
\begin{eqnarray}
G^{tt} &=& 1 \cr
G^{\gvfi\gvfi} &=& \frac{s^2}{c^2} - \frac{4}{k} \cr
G^{3\gvfi} &=& \frac{s^2}{2 c} x_0  \cr
G^{22} &=& ( \frac{c-1}{c+1} - \frac{1}{k-2} ) x_0{}^2
          +  ( \frac{c+1}{c-1} - \frac{1}{k-2} ) x_3{}^2 \cr
G^{33} &=& ( \frac{s^2}{4 c^2} - \frac{1}{k-2} ) x_0{}^2
          +  ( \frac{c+1}{c-1} - \frac{1}{k-2} ) x_2{}^2 \cr
G^{23} &=& - ( \frac{c+1}{c-1} - \frac{1}{k-2} ) x_2 x_3
\end{eqnarray}
}
\begin{eqnarray}
\gw^1 &=& ( \frac{s^2}{c^2} - b)^{-\frac{1}{2}}\, d\gvfi \cr
\gw^2 &=& \frac{c+1}{s\,\gr} \, \gb^{\frac{1}{2}}\, ( dx_0 +
\frac{f}{1 - b\,\frac{c^2}{s^2}}\, \frac{x_3}{2} \, d\gvfi - (f-1)
\frac{x_3}{x_0} \, dx_3 ) \cr
\gw^3 &=& \frac{c-1}{s\,\gr} \, \left( \frac{f}{1- a
\frac{c-1}{c+1}} \right)^{\frac{1}{2}}\,
( dx_3 - (1 - b\,\frac{c^2}{s^2} )^{-1}\, \frac{x_0}{2} \, d\gvfi ) \cr
\gw^4 &=& dt
\end{eqnarray}
and after solving the differential equations, the dilaton
\be
D = D_0 + \ln \frac{s^3 \; c}{|\det \, G |^{\frac{1}{2}} }
\ee
where
\begin{eqnarray}
det\, G &=& \frac{\gb \; f}{(1 - a\, \frac{c-1}{c+1})
(\frac{s^2}{c^2} - b) \, \gr^4 } \cr
\gb^{-1} &=& 1 - \frac{c+1}{c-1} \left( a + ( f - 1) (
\frac{c+1}{c-1} - a) \,\frac{x_3{}^2}{x_0{}^2} \right) \cr
f^{-1} &=& 1 - \frac{a\, b}{\ge} \; (\frac{c+1}{c-1} - a)^{-1}\;
\frac{ (1-\ge) c^2 -1}{(1- b) c^2 -1}\; \frac{x_0{}^2}{\gr^2}
\end{eqnarray}
and $\; a=\frac{1}{k-2},\; b=\frac{4}{k},\; \ge = \frac{2}{k-1}\,$ .

As usual the exact results are not very enlightening and in general the
singularity structure becomes highly complicated.
Also regions of different signature appears, fact related to the signs in
the arguments of the square roots in (6.12), giving rise to bizarre
geometries and possible topologies.
For example, for $\, 0<k<2\, $ is easy to see that the signature is
strictely minkowskian (within the natural range of the group parameters)
with $\gvfi$ being the time like coordinate.
However some interesting interpretations can be given.

\n {\bf The black plane metrics}

Let us consider metrics of the form
\be
G_{bp} = -f(x) \; d\gt^2 + f(x)^{-1}\; dx^2 + dy^2 + dz^2
\ee
Obviously the topology is $P\times Q$ where $P$ is a plane (or some
compactified version of it) and $Q$ an indefined signature submanifold
parametrized by $(\gt ,x)$ coordinates where the geometry is characterized by
the function $f$.

Let us first analyze a ``regular" case with
\be
f_r (x) = 1 - \frac{\cosh^2 a x_h}{\cosh^2 a x}
\ee
where $a, x_h$ are positive real constants, and introduce the
``distorted" coordinate
\be
x_* = x \; +\; \frac{1}{2\, a\, \tanh a x_h}\; \ln\frac
{\sinh a|x-x_h|}{\sinh a|x+x_h |}
\ee
The inverse relation $x(x_*)$  distinguishes  three patches:
I for $x>x_h$, II for $|x|<x_h$ and III for $x<-x_h$.
By defining null coordinates $ u = \gt + x_* \; , \; v = \gt - x_* $ in
regions I and III, and $ u = x_* + \gt \; , \; v = x_* - \gt $ in region II,
the  metric takes the general form
\be
G_{rbp} = - |f_r (x)|\; du \; dv + dy^2 + dz^2
\ee
The metric is regular in all three patches as can be seen from the scalar
curvature (that characterizes all the curvature tensor)
\be
R = 2 \; a^2 \; \cosh^2 a x_h \;\; \frac{-3 + 2 \cosh^2 a x}{\cosh^4 a x}
\ee
Then we can glue them as is usually done and the maximally extended conformal
Penrose diagram for $Q$ (where each point represents $P$) so obtained is
similar to that of the Kerr solution of general relativity
(for $M^2 > a^2, \theta = 0 $)
\footnote{See for example figure 27 in page 312 of reference [20].}
with $ r_\pm \sim \pm x_h $, and the manifold described by it is geodesically
complete.
Clearly $x\rightarrow \pm \infty$ are asymptotically flat regions, and
$x = \pm x_h$ are horizons for observers there (in regions I/III);
the geometry is then naturelly interpreted as a ``regular black plane"
hidden in region II.
Its Hawking temperature can be computed by standard methods [4]
\be
T_r = \frac{a}{2\pi} \tanh a x_h
\ee

Let us consider now a ``singular" case defined by
\be
f_s (x) = 1 - \frac{\sinh^2 a x_h}{\sinh^2 a x}
\ee
The distorted coordinate is now defined as in (6.17) with the replacement
$$
\tanh a x_h \rightarrow ({\tanh a x_h})^{-1}
$$
But now the curvature is
\be
R = 2 \; a^2\; \sinh^2 a x_h \;\; \frac{3 + 2 \sinh^2 a x}{\sinh^4 a x}
\ee
that togheter with (6.21) reveals the existence of flat regions for
$|x|\rightarrow \infty$, but also displays a true singularity at $x=0$.
Due to this crucial fact we can follow the standard procedure as before
and write $G_{sbp}$ as in (6.18), but now we can only glue region I with
``half" region II (until the singularity, remember that here $x$ is timelike)
because we can not go beyond the singularity where analyticity breaks down;
similar remarks are made for regions III and the other half of region II,
which are ``parity" reflected patches of the first ones.
The maximally extended conformal Penrose diagram is then similar to
Schwarzchild's.
We can say that the singularity at $x=0$ separates two worlds;
we can not certainly pass through the singular black plane, and once we
go across the horizon at $x_h$ we will die there after finite proper time.
The Hawking temperature for this ``singular black plane" is
\be
T_s = \frac{a}{2\pi} \coth a x_h
\ee

Now let us establish what these geometries has to do with us.
Let us consider the general case of finite $k \neq 2,3,4$.
Then it is not difficult to show that exists $0<t_k <\infty$ such that the
exact solution given by (6.12,13) has the limit
\be
G\;\stackrel{t\gg t_k }{\longrightarrow}\;  dt^2 + \frac{k}{k-4}\;
d\gvfi^2 + \frac{k-2}{k-3}\; \frac{dr^2}{1- r^2} +
\frac{k-4}{k-3}\; \frac{r^2}{\frac{k-4}{k-2} - r^2} \; d\psi '_P{}^2
\ee
\be
D\;\stackrel{t\gg t_k }{\longrightarrow}\; 4\; t + \frac{1}{2} \ln |(1 -
r^2 ) (\frac{k-4}{k-2} - r^2)|
\ee
where polar coordinates ($r\equiv\sin R ,\psi $) as in (4.12) has been
introduced and
$$
\psi '_P = \psi - \frac{k}{k-4}\;\; \frac{\gvfi}{2}
$$
Now let us take $0<k<2$ (for example, the conformal value $k=k_{-}$
discussed after (5.9)).
Then by making the change of variables
\be
r^2 = 1 - \frac{2}{2-k} \; \sinh^2 a x
\ee
with $a = \frac{1}{\sqrt{|k-2|}}$, it is easy to show that the
$line$ $element$
$$
ds^2 = (k-3)\; G
$$
tends to the the regular black plane metric with the further identifications
\begin{eqnarray}
  y &=& i \;\sqrt{|3-k|}\;\; t  \cr
  z &=& \sqrt{k\; \frac{|3-k|}{|4-k|}} \;\; \gvfi \cr
\gt &=& i \;\sqrt{|4-k|} \;\; \psi '_P
\end{eqnarray}
and $x_h$ defined by $\;  \sinh^2 a x_h = 1 - k/2 $.

On the other hand, in the case $4<k<\infty $ the change of variables
\be
r^2 = 1 - \frac{2}{k-2} \; \cosh^2 a x
\ee
leads to
\be
ds^2 \; \stackrel{t\gg t_k }{\longrightarrow}\; - G_{sbp}
\ee
with $a$ as before, $ \sinh^2 a x_h = -2 + k/2 $, and the identifications
are (6.27) with the replacement $z\rightarrow i z $.
The dilaton field in both cases is given by
\be
D\;\stackrel{t\gg t_k}{\longrightarrow}\; 4\; t + \ln \sinh 2 a |x|
\ee

{}From these results we are in conditions of interpreting the $exact$
solutions (6.12), as we made in the $k=\infty$ case, as some kind of
instantons
that ``tunnel" from  $t\rightarrow 0$ highly singular universes (whose
expressions being little ilumining we do not write) to static black plane like
universes for $t\gg t_k $.
We also notice from (6.30) that $t\gg t_k $ is a weak coupling phase except
near the black plane $x \rightarrow 0 $ where we go to an strong coupling
region.

Let us finally remark that the $\gK$ field introduced in (6.6) results
$k$-independent, as verified for some models in [5].
This result gives further strong support to the non renormalization theorem
conjectured there for any GWZM from path integral measure conformal invariance
arguments.
\newpage


\section{The dual backgrounds}

In reference [21] was showed that it is possible to obtain another solution
to the one loop equations (2.2) starting from one which has an isometry.
Explicitely, if $(G, B, D)$ are backgrounds satisfying (2.2) that in some
coordinate system are independent of the coordinate $\gvfi$, then
\begin{eqnarray}
\tG_{\gvfi\gvfi} &=& G_{\gvfi\gvfi}{}^{-1} \cr
\tG_{\gvfi\ga} &=& \frac{B_{\gvfi\ga}}{G_{\gvfi\gvfi}} \cr
\tG_{\ga\gb} &=& G_{\ga\gb} - \frac{1}{G_{\gvfi\gvfi}}
( G_{\gvfi\ga}\; G_{\gvfi\gb} \; -\; B_{\gvfi\ga}\; B_{\gvfi\gb} ) \cr
\tB_{\gvfi\ga} &=& \frac{G_{\gvfi\ga}}{G_{\gvfi\gvfi}} \cr
\tB_{\ga\gb} &=& B_{\ga\gb} + \frac{1}{G_{\gvfi\gvfi}}
( G_{\gvfi\ga}\; B_{\gvfi\gb} \; -\; B_{\gvfi\ga}\; G_{\gvfi\gb} ) \cr
\tD &=& D + \ln |G_{\gvfi\gvfi}|
\end{eqnarray}
where $\ga,\gb \ne \gvfi$, is also a solution.
The existence of it is sometimes referred as ``target space duality" or
``abelian duality".
As we saw in Section 4, (4.10,11) fulfills the requirements and then
a dual solution may be straightforwardly obtained from (7.1).
For sake of completeness we present it,
\begin{eqnarray} \tG &=& dt^2 \; +\; \frac{1}{G_{\gvfi\gvfi}} \; d\gvfi ^2
+ \frac{1}{4\; G_{\gvfi\gvfi} \; \gr ^2} \; (\;\;
( \frac{4\; c^2}{(c-1)^2} + \frac{x_0{}^2}{\gr ^2} )\;  dx_0{}^2
+ 2\;\; \frac{x_0\, x_3}{\gr ^2}\;\; dx_0 \,dx_3 \cr
&+& (\, \frac{4\; c^2}{(c+1)^2} + \frac{x_3{}^2}{\gr ^2}\,)\; d x_3{}^2
\; )\cr
\tB &=& \frac{1}{2 \; G_{\gvfi\gvfi}\; \gr ^2} d\gvfi \wedge (
\frac{c+1}{c-1}\; x_3 \; dx_0 \;- \; \frac{c-1}{c+1} \; x_0 \; dx_3 ) \cr
\tD &=& \tD_0 + \ln |s^4 \; \gr ^2 \; G_{\gvfi\gvfi} |
\end{eqnarray}
We notice that the crossing terms in (4.3) does not appear in (7.2), at
expenses of the axionic field.
Also the metric present a submetric in the $(t,x_0,x_3 )$ variables;
formally the Cotton-Darboux theorem [20] assures us that it is possible to
diagonalize it but unfortunately we have not succeeded in doing it.

In [22] was showed that if the coordinate $\gvfi$ is periodic, then both
solutions are equivalent, i.e., they describe the same conformal theory.
In the natural range of our parameters, $\gvfi$ is in fact periodic, and
then both (4.10,11) and (7.2) should be equivalent.
This can be understood from the GWZM point of view by noting that, having
gauged a subgroup with a semisimple algebra containing a $u(1)$ subalgebra,
there exists the possibility of considering other model by $axial$
gauging the $u(1)$ (see footnote 2).
We then conclude that the one loop backgrounds (7.2) are those of the
$SU(2,1)/SU(2)_{vector}\times U(1)_{axial} \; $ GWZM.

\newpage


\section{Conclusions}

We have presented in this paper an study of the possible effective
geometries underlying a coset model based on the pseudo-unitary group
$SU(2,1)$, to our knowledge the first one that considers $SU(p,q)$ groups
with $p+q>2$.

In the natural range of the parameters the one loop metric is strictely
positive definite and so it does not present ``horizons", but is
singular on two dimensional manifolds $t=0$ (disk) and $\gr =0$ ($\Re^2$).
It may be possible that by changing the topology (e.g., limiting the range of
coordinates or compactifying some dimensions) a ``regular" gravitational
instanton may be obtained.
For example, if we introduce in (4.13) the $x$ variable by
\be
\sin \, R = e^{- x + t^\gn} \;,\;\; 0<\gn<1
\ee
then we have for $t\gg 1\,$,
\be
G \rightarrow dt^2 + d\gvfi ^2 - dx^2 - d\psi_P{}^2
\ee
that is, $G$ results asymptotically flat on $\Re^2\times T^2$ (the
Riemann tensor in fact vanishes).
Anyway it does not seem any such modified theory will be fully represented
by an exact conformal
field theory, because only some patch would be covered by the GWZM
considered here.

For finite $k$ (the physical case) the picture drastically changes.
Regions of different signature appears, and the structure of the
singularities becomes highly complicated.
In the examples considered we remain with them, differing from the
$2-d$ black hole model where a possible mechanism to evite the
singularity seems to work [23].

A question non addressed in this paper is the global topology of the
exact target manifold; we have in fact loosely ignored the ranges of the
coordinates in the discussions of section 6, although is clear that
(6.12,13) is presumibly a solution of the (unknown) exact background
field equations independently of them.
In our opinion only the study of the quantum theory of the model and possible
consistency conditions (e.g., identification of field operators with current
algebra primary fields, renormalization, unitarity, ecc.) needed for its
existence can give light on the problem.

Finally we remark that, as it occurs with other string solutions, the
existence of event horizons with topology different from $S^2$ (in our
case, a plane) is not in contradiction with Hawking's theorem, because
our solution has $\gL =12 >0$ that gives a negative Liouville potential
in (5.10) which violates the dominant energy condition [4].

\newpage


\appendix
\n{\large {\bf Appendix} }

\section{U(p,q) parametrization}
Let $g$ an arbitrary element of ${\CC}^{(p+q)\times (p+q)}$,
\be
g = \left (\matrix{        A    &B' \cr
                     C'^{\dag}  &D \cr} \right)
\ee
where $A$ is $p\times p$, $D$ is $q\times q$ and $B'$, $C'$ are $p\times q$
complex matrices.
Let $\eta$ the diagonal element given by $A=1$, $D=-1$ and $B'=C'=0$.
Then the condition $ g\eta g^\dagger = \eta$ define the elements of $U(p,q)$,
and leads to the set of equations
\begin{eqnarray}
A A^{\dag} & = & 1 + B' B'^{\dag}  \cr
D D^{\dag} & = & 1 + C'^{\dag} C'  \cr
       A C'& = & B' D^\dag
\end{eqnarray}
The first two equations are solved respectively by
\footnote{For any complex $M$ the matrix $M M^\dag$ is certainly hermitic
and non-negative, and then arbitrary powers of it are well defined
through its diagonal form.}
\begin{eqnarray}  A&=&(1 + B'B'^{\dag}
)^{\frac{1}{2}}\; U \cr
                  D&=&(1 + C'^{\dag} C')^{\frac{1}{2}}\; V
\end{eqnarray}
where $U \in U(p)$ and $V \in U(q)$ are arbitrary.
If we reparametrize: $C'=U^{\dag} C, \, B'=B V$, last equation in (A.2)
is solved by  $B=C$, and then
\begin{eqnarray}
 g(C,U,V) &=& T(C)\; H(U,V) \cr
     T(C) &=& \left( \matrix{
               (1 + C C^{\dag})^{\frac{1}{2}}   &C\cr
                          C^{\dag}              &(1 + C^{\dag}C)^{\frac{1}{2}}
\cr} \right) \;\; \in U(p,q)/U(p)\otimes U(q) \cr
   H(U,V) &=& \left( \matrix{
                          U   &0\cr
                          0   &V\cr } \right) \;\; \in U(p)\otimes U(q)
\end{eqnarray}
By making the change of variables
$ C = (N N^\dag)^{-{1\over2}} \sinh (N N{^\dag})^{1\over 2} N $
we can write the coset element as
\footnote{For an extensive treatment of coset spaces, see [24].}
\be
T(C) = exp \left( \matrix{ 0          &N \cr
                           N^{\dag}   &0 \cr } \right)
\ee
Let us remark that analogous coset decompositions can be considered in terms
of  non compact versions of the maximal compact subgroup $U(p)\otimes U(q)$.
Also they lead to the corresponding ones to the group $O(p,q)$ by taking
the apropiate real sections.

Under an adjoint transformation $g^h = h g h^\dag\,$
with  $ h \equiv H(h_1,h_2^\dag ) \in U(p)\otimes U(q)$, $g$ transforms as:
\begin{eqnarray}
C^h &=& h_1 \; C \; h_2 \cr
U^h &=& h_1 \; U \; h_1^\dag \cr
V^h &=& h_2 \; V \; h_2^\dag
\end{eqnarray}

For $q=1$, $C$ is a p-dimensional complex vector and $V$ a phase.
By restricting ourselves to $SU(p,1)$ we have $u \equiv \det U = V^\dag$;
clearly the topology of $SU(p,1)$ is ${\Re}^{2p} \times U(p)$ and its maximal
compact subgroup is $U(p)$.
A coset decomposition of $U(p)$ wrt its invariant subgroup $SU(p)$ yields,
for $p=2$, the parametrization used in the text (c.f. (3.3)), $U(2)$ being
generated by $\{\gl_1, \gl_2, \gl_3, \gl_8\}$ and the argument of $T$ in
(A.5) by the other Gell-Mann matrices.

\bigskip

\section{Some relations for $4\times 4$ matrices}

Here we collect some useful formulae for computing $\gl^{-1}$ in Section 3.

Given an arbitrary  $4\times 4$ matrix in the form
\be
\gl = \left(\matrix{
                       M     &\bf{m_1}  \cr
                       \bf{m_2}^t   &m_0  \cr}\right)
\ee
where $M$ is a $3\times 3$ matrix, $\bf{m_1}, \bf{m_2}$ are 3-vectors and
$m_0$ a number, then direct inspection shows that its cofactor matrix is
given by
$$
\gl^{c} = \left(\matrix{
                          {\tilde M}       &-M^c \bf{m_2} \cr
                           -\bf{m_1}^t M^c      &m        \cr}\right)
$$
\be
{\tilde M} = m_0 M^c - (M^t - trM\, 1) (\bf{m_2} \bf{m_1}^t - \bf{m_1}^t
\bf{m_2} \; 1) - (\bf{m_2} \bf{m_1}^t M^t - \bf{m_2}^t M \bf{m_1}\; 1)
\ee
and its determinant by
\be
l \equiv \det\gl = m_0 \; m - \bf{m_1}^t M^c \bf{m_2}
\ee
In these equations $m\equiv \det M$, and
\begin{eqnarray}
M^c &=& (M^2)^t - trM M^t + tr M^c 1 \cr
2\; tr M^c &=& (trM)^2 - trM^2
\end{eqnarray}
For $M\equiv 1-A$ we have
\begin{eqnarray}
m &=& 1 - a + tr(A^c -A), \;\;\;\; a\equiv \det A \cr
M^c &=& (1-trA)\, 1 + A^t + A^c
\end{eqnarray}
Finally, from (B.5) for $M\equiv R \in O(3)$ we have the useful relation
\be
R^2 - trR \, R = R^t -trR \,1
\ee

\bigskip


\section{$SU(2)$ miscelaneous.}

An arbitrary matrix $X \in SU(2)$ can be written as
\be
X =  x_0 \, 1 + i \, \bf{x}\cdot \bf{\gs} = \left( \matrix{  z & w \cr
                                       -w^* & z^* \cr } \right)
\ee
where $\bf{\gs}$ are the Pauli matrices, $tr(\gs_i \gs_j) = 2
\,\delta_{ij}$, and
\begin{eqnarray}
z&=&x_0 +i x_3 \, , \;\; w= x_2 + i x_1  \cr
1&=& x_0{}^2 + {\bf x} \cdot {\bf x} = z z^* + w w^*
\end{eqnarray}
\ignore{
\n In the usual $S^3$ parametrization
($\; \gq_1, \gq_2 \, \in [0,\pi] \;\;,\;\;\gq \, \in [0,2\pi) \; $)
\begin{eqnarray}
x_0 &=& \cos \gq_1 \cr
x_1 &=& \sin \gq_1 \sin \gq_2 \sin\gq \cr
x_2 &=& \sin \gq_1 \sin \gq_2 \cos\gq \cr
x_3 &=& \sin \gq_1 \cos \gq_2
\end{eqnarray}
}
If $\, w = \gr\, e^{i \theta}\, ,\, 0<\gr <1 \, ,\,$ we have
\ignore{
\be
X = e^{ i\frac{\gq}{2} \gs_3 }
    e^{ i\frac{\gq_2}{2} \gs_1}
    e^{ i \gq_1 \gs_3}
    e^{-i\frac{\gq_2}{2} \gs_1}
    e^{-i\frac{\gq}{2} \gs_3}
}
\begin{eqnarray}
X &=& e^{ i\frac{\gq}{2} \gs_3 }\, {\overline X}\,
e^{-i\frac{\gq}{2}\gs_3} \cr
{\overline X} &=& \left( \matrix{  z   & \gr  \cr
                                  -\gr &  z^* \cr}\right)
\end{eqnarray}
The Maurer-Cartan form associated with $X$ is defined by
\begin{eqnarray}
\gw(X) &\equiv& X^{-1}dX = i \; \bf{U}(\bf{x}) \cdot \bf{\gs} \cr
U_i (x) &=& x_0 \, dx_i - x_i \, dx_0 + \ge_{ijk} \, x_j \, d x_k
\end{eqnarray}
and analoguously $\;{\overline \gw}(X) \equiv dXX^{-1} = i \;
\bf{\overline U}(\bf{x})\cdot \bf{\gs} $, with $\bf{\overline
U}(\bf{x})=-U(-\bf{x})$.

The adjoint representation matrix of $X$,
\be
 R(X)_{ij} \equiv \frac{1}{2} tr( \gs_i X \gs_j X^\dag )
\ee
\n is given by
\be
R(X)_{ij} = (2\, x_0^2 - 1) \gd_{ij} + 2 \,( x_i\, x_j + x_0\, \ge_{ijk}\,
x_x)
\ee
Particularly useful in the text are the variables
\begin{eqnarray}
R_{33} &=& 1 - 2 \, (x_1{}^2 + x_2{}^2) = -1 + 2 (x_0{}^2 + x_3{}^2) \cr
   trR &=& 4 x_0{}^2 - 1
\end{eqnarray}
{}From these formulae the following expressions are obtained:
\newpage
\begin{eqnarray}
\bf{U}\cdot \wedge *\bf{U} &=& dx_0\wedge * dx_0 + d\bf{x}\cdot \wedge *
d\bf{x} \cr
2\, \bf{U} \cdot \wedge *\bf{\overline U} &=& (trR + 3) dx_0{}
\wedge *dx_0 + (trR - 1) d\bf{x}\cdot \wedge *d\bf{x} \cr
\bf{U}\cdot \wedge \bf{\overline U}  &=& 2\,x_0 \; \ge_{ijk} \; x_i \;
dx_j\wedge dx_k\cr
U_3 \wedge {\overline U}_3&=& (1- R_{33})\, (x_0 \; dx_3-x_3 \;dx_0)
\wedge d\gq \cr
\bf{e_3}\cdot R\; \bf{\overline U} &=& -2\, d(x_0 x_3 ) + trR\;
\big( x_0 \; dx_3 - x_3 \; dx_0  + \frac{1}{2} ( 1 - R_{33}) \; d\gq \big)\cr
\bf{e_3}\cdot R^t\;\bf{U} &=& -2\, d(x_0 x_3 ) + trR\;
\big( x_0 \; dx_3 - x_3 \; dx_0  - \frac{1}{2} ( 1 - R_{33} ) \; d\gq \big)
\end{eqnarray}

\newpage


\section*{References}
\begin{enumerate}
\it E. Witten, \prd 44 (1991), 314;\\
G. Mandal, A. Sengupta and S. Wadia, Mod. Phys. Lett. {\bf A}6 (1991), 1685.
\it C. Callan, D. Friedan, E. Martinec and M. Perry, \np 262 (1985), 593.
\it D. Karabali: ``Gauged WZW models and the coset construction of CFT",
Brandeis preprint BRX TH-275, July 1989, and references therein;\\
S. Chung and S. Tye: ``Chiral gauged WZW theories and coset models in CFT",
Cornell preprint CLNS 91/1127, January 1992.
\it G. Horowitz, ``The dark side of String theory: black holes and black
strings", proceedings of the 1992 Trieste Spring School on String theory and
Quantum gravity, and references therein.
\it I. Bars and K. Sfetsos, USC-92/HEP-B1, B2 and B3 preprints (1992)
\it R. Dijkgraaf, E. Verlinde and H. Verlinde, \np 371 (1992), 269.
\it D. Gershon: ``Exact solutions of four dimensional black holes in string
theory", TAUP-1937-91, December 1991.
\it E. Kiritsis, Mod. Phys. Lett. {\bf A}6 (1991), 2871.
\it P. Ginsparg and F. Quevedo, \np 385 (1992), 527.
\it M. Green, J. Schwarz and E. Witten: ``Superstring theory", vol. 1,
Cambridge University Press, Cambridge (1987).
\it R. Wald, ``General Relativity", University of Chicago Press, Chicago
(1984).
\it P. Goddard and D. Olive, \ijmp 1 (1986), 303, and references therein.
\it F. Quevedo: ``Abelian and non-abelian dualities in string
backgrounds", Neuch\^{a}tel preprint NEIP93-003, July 1993.
\it E. Alvarez, L. Alvarez-Gaum\'{e} and Y. Lozano: ``On non abelian
duality",  CERN preprint, March 1994.
\it A. Sen: ``Strong-weak coupling duality in 4-d string theory", Tata
Institute preprint, hep-th/9402002, September 1994.
\it G. Gibbons and S. Hawking, \cmp\, 66 (1979), 291.
\it K. Gawedzki: ``Non compact WZM CFT", I.H.E.S. preprint (1991).
\it T. Eguchi, P. Gilkey and A. Hanson, Phys. Rep. 66 (1980), 213.
\it D. L\"{u}st: ``Cosmological string backgrounds", CERN preprint, March
1993.
\it S. Chandrasekhar, ``The mathematical theory of black holes", Oxford
University Press, New York (1983).
\it T. Buscher, \pl 194 (1987), 59, \pl 201 (1988), 466.
\it M. Ro\v{c}ek and E. Verlinde, \np 373 (1992), 630.
\it M. Perry and E. Teo: ``Non singularity of the exact two dimensional
black hole", DAMPT preprint hep-th/9302037.
\it R. Gilmore: ``Lie groups, Lie algebras and some of their
applications", Wiley, New York (1974).
\ignore{
\it G. Moore and N. Seiberg, \pl 212 (1988), 451.
\it C. Callan, J. Harvey and A. Strominger, \np 359 (1991), 611.
\it C. Callan, R. Myers and M. Perry, \np 311 (1988), 673.
\it C. Callan, S.Giddings, J. Harvey and A. Strominger, \prd 45 (1992), 1005.
\it E. Verlinde and H. Verlinde, ``A unitary S-matrix for 2D black hole
formation and evaporation", PUPT-1380/IASSNS-HEP-93/8 preprint, february 1993.
}

\end{enumerate}

\end{document}